\documentclass[fleqn,twoside]{article}
\usepackage{espcrc2}

\usepackage{graphicx}
\usepackage[figuresright]{rotating}

\newcommand{\AmS}{{\protect\the\textfont2
  A\kern-.1667em\lower.5ex\hbox{M}\kern-.125emS}}

\title{Pion Rescattering in Nuclei}

\author{E.A.\ Paschos\address[UniDo]{Institute for Physics, University of Dortmund,\\
              44221 Dortmund, Germany},
        I.\ Schienbein\address{Deutsches Elektronen Synchroton (DESY),\\
              22603 Hamburg, Germany},
        J.-Y.\ Yu\addressmark[UniDo].\\
       (presented by E.A.\ Paschos)}
  
\begin{document}
\begin{abstract}
Nuclear corrections are presented for neutrino and electron induced reactions
in a pedagogical manner. The formalism is demonstrated with
numerical studies and is shown to produce substantial corrections
in channels where the pions have the same charge with the exchanged current. 
Two comparisons with available data show consistency of the model.
Additional experimental results along these lines will improve the accuracy of
the predictions and enhance the discovery potential of experiments.
\vspace{1pc}
\end{abstract}

\maketitle

\section{Formulation of the Model}

It is widely recognized that nuclear corrections in neutrino reactions
play an important role because they alter sometimes the neutrino--nucleon
interaction at the 10-30\% level. There are standard questions that are
frequently asked on this topic for which the answers are not yet
confirmed by observations. Among them I will mention three.
\begin{itemize}
\item[(i)] How big is the absorption of hadrons and in particular
of pions within a nucleus? 
\item[(ii)] How does nuclear matter affect the propagation of 
resonances themselves? 
\item[(iii)] When a resonance decays within a nucleus, what are the
changes in the energy and angular distribution of the decay products,
due to interactions within the medium?
\end{itemize}
Such questions are important for the interpretation of the new
experiments and the precise determination of neutrino parameters
which occur in neutrino oscillations and the discovery of CP violation
in the leptonic sector \cite{ref1}.

This article tries to present a consistent and easy to follow summary
of the calculations in the ANP model 
\cite{ref2} 
and then mentions a few comparisons with data which are now possible. 
First I will discuss a general framework within which nuclear 
effects are calculated.

Resonances like the $P_{33},\, P_{11},\, S_{11}$ and $D_{13}$ are
produced within nuclei with the cross sections proportional
to the density of nuclear matter.  They either propagate within
the nuclear medium or decay producing hadrons like pions, a 
proton or a neutron, which propagate in the medium before they exit
from the nucleus.  In this article we assume that the resonances
either decay immediately or are absorbed.  The latter means that 
they interact with the medium producing excited nuclear states
which subsequently vibrate decaying eventually to the ground state.
The extra energy in this case is transferred to thermal or other
unobserved energy.  The absorption of pions will be parametrized
by a phenomenological function $\sigma_{abs}(W)$, with $W$ the 
invariant mass of the pion nucleon system \cite{ref9}.

In this article we shall consider the pions produced from the 
decays of the resonances and the interaction of the pions with an
isoscalar medium.  The propagation of the pions is characterized
by three eigenfunctions with characteristic eigenvalues $\lambda_i$: 
\begin{eqnarray}
\lambda_1 = 1, & \quad\quad & q_1=\left( 
\begin{array}{c} 
1 \\ 1 \\ 1 
\end{array}\right),\nonumber\\
\lambda_2 = \frac{5}{6}, &\quad\quad & q_2 =\left(
\begin{array}{c} 
1 \\ 0 \\ -1
\end{array}\right),\nonumber\\
\lambda_3 = \frac{1}{2}, &\quad\quad & q_3=\left(
\begin{array}{c} 
1 \\ -2 \\ 1
\end{array}\right) \ .
\end{eqnarray}
The eigenfunctions $q_i,\, i=1,\,\ldots,\, 3,$ denote the pion charge
multiplicities (populations) after successive scatterings.  More
precisely the pion admixtures included in the eigenfunctions are
reproducing themselves after each scattering.
Assuming charge--symmetry for the interactions there are three 
independent transport functions.  In addition, assuming 
$\Delta$--resonance dominance the elementary interactions are written
in terms of the $\pi^+p$ cross section.  The probability of a 
pion origninally at point $y$ to propagate to point $x$ and
interact there is governed by an exponential law
\begin{equation}
P(x,y) = e^{-K(x-y)}K(y)
\end{equation} 
with $K=\rho(x)\sigma_g(W)$ and $\sigma_g(W)$ an effective 
cross section given by
\begin{eqnarray}
\left( 
\begin{array}{c} 
\sigma_g(\pi^+) \\ 
\sigma_g(\pi^0) \\
\sigma_g(\pi^-)
\end{array}\right)
&  = &
\left( 
\begin{array}{ccc} 
1 & 0 & 0 \\ 
0 & 1 & 0 \\ 
0 & 0 & 1
\end{array}
\right) \sigma_{abs} \nonumber\\
&+ & \left( 
\begin{array}{ccc} 
\frac{5}{9} & \frac{1}{9} & 0\\
\frac{1}{9} & \frac{4}{9} & \frac{1}{9} \\
0 & \frac{1}{9} & \frac{5}{9}
\end{array}\right)\frac{3}{2}\sigma_{\pi^+ p}\ . \label{eq3}
\end{eqnarray}
In this form, the effective cross section depends on the absorption which is 
assumed to be the same for the three types of pions plus an
interaction cross section which includes charge--exchange terms.
The transport problem is a matrix problem which is solved as an
eigenvalue problem.  The absorption term, being proportional to the 
unit matrix, brings an overall factor
\begin{equation}
A=g(Q^2,W)\, f(\lambda=1,\,W)
\end{equation}
consisting of a Pauli suppression factor $g(Q^2,W)$ at the 
production point times a transport function $f(\lambda=1,\,W)$
for the eigenvalue $\lambda=1$.  The eigenvalues and eigenfunctions
in eq.~(1) correspond to the interaction matrix, i.e.\ the second matrix in 
eq.~(\ref{eq3}).  
As we describe later
on, each eigenvalue has its own transport factor $f(\lambda,\ W)$.

The absorption cross section is a phenomenological function which
contains several nuclear effects; for example excitation of the 
nucleus, a short propagation of the $\Delta$--resonance before it
decays, etc.  It is determined phenomenologically from data.

Let us denote by $d\sigma^i$ with $i=+,\, 0,\, -$ the 
neutrino--nucleon cross section averaged over protons and neutrons
at production point within the nucleus.  Similarly denote by
$d\sigma(T^A, \,\pi^i)$ the number of pions with charges
$i=+,\, 0,\,-$ emerging from the nucleus $T^A$.  The isospin
symmetry of the problem gives the following relation
\begin{eqnarray}
\left[
\begin{array}{l}
d\sigma(T^A;\, \pi^+)\\
d\sigma(T^A;\, \pi^0) \\ 
d\sigma(T^A;\, \pi^-)
\end{array}\right]
= A(Q^2,\, W) \,\,
M
\left[ 
\begin{array}{c} 
d\sigma(\pi^+)\\ 
d\sigma(\pi^0)\\
d\sigma(\pi^-)
\end{array}\right]
\end{eqnarray} 
with
\begin{equation}
M=
\left[ 
\begin{array}{ccc} 
1-c-d & d & c\\ 
d &  1-2d & d\\
c & d & 1-c-d
\end{array}\right]\, .
\end{equation}
The parameters $c$ and $d$ depend on the charge exchange and are
given in ref.\ \cite{ref2}.  
They will be defined also below.  Taking the sum of
all pions we arrive at the relation
\begin{equation}
d\sigma(T^A;\, \pi^I) = A(Q^2,\, W)
   \!\!\!\!\!\!\sum_{K=+,0,-} M_{IK}d\sigma(\pi^K)\ .
\end{equation}
When the Pauli factor is weighted by $\frac{d\sigma}{dQ^2}$, it
gives a value between 0.96 and 0.98 and consequently $A(W)$
gives practically the suppression introduced by absorption.  
This is seen in table \ref{tab1} where the Pauli factor is given as a 
function of $Q^2$ in GeV$^2$.

\renewcommand{\arraystretch}{1.2} 
\begin{center}
\begin{tabular}{|c|c|}
\hline
$Q^2$ & $g(Q^2,\, M_{\Delta})$\\
\hline
0.05 & 0.87 \\
0.20 & 0.98 \\
0.40 & 1.00 \\
\hline
\end{tabular} \label{tab1}
\end{center}
\begin{center}
Table 1: Pauli factor in dependence of $Q^2$.
\end{center}

\noindent We describe next the transport functions.\\

In general the transport problem is a random--walk problem of the  pions
within the nucleus, which is usually included in Monte--Carlo
programs as a multi--scattering process.  In ref.\ \cite{ref2}
there are analytic solutions for the general case as well as
approximate solutions for specific cases.  

Since the pion--nucleon
cross section in the $\Delta$--region is peaked in the 
forward--backward direction 
($\sigma(\theta) \simeq 1+3\cos^2\theta$) we shall consider an
average of the cross section in the forward and backward
hemispheres.  We define a second Pauli factor $h_{\pm}(W)$ 
corresponding to the scattering of the pion on a bound nucleon.
The Pauli factors $h_{\pm}(W, \cos\theta)$ are given in Appendix
C, part 2 (page 2142) of ref.~\cite{ref2}.
At the same time we introduce an average over forward and 
backward hemispheres
\begin{eqnarray}
\langle \sigma_{\pi^+ p}(W)h_+(W)\rangle & = & 
\int_{1}^{0} \sigma_{\pi^+p}\,(W,\, \cos\theta)\nonumber\\
&\times& h_+(W,\, \cos\theta)\, d\cos\theta\, 
\end{eqnarray}
with a similar integration defining the averaging over  $h_-(W)$. Then 
the probability
for transmission in the forward direction is given by
\begin{equation}
f_{+}(\lambda, \, L,\, W) =
\frac{1}{L(b)}\int_0^{L(b)} dy\, e^{-yK(1-\lambda\mu_{+})}
\end{equation}
with
\begin{eqnarray}
yK(1-\lambda {\mu_{+}}) &=& y\rho(0)
\big[\sigma_{abs}(W)\nonumber\\
& & -\frac{1}{3}\lambda \langle \sigma_{\pi^+ p}(W)\, h_{+}(W)\rangle\big]\, , 
\end{eqnarray}
$\lambda$ one of the three eigenvalues, and $L(b)$ the effective profile
of the nucleus at a given impact parameter $b$.  In other words,
when the neutrino produces a pion at an impact parameter $b$,
the pions begin to rescatter forward along the line
of the impact parameter until they exit from the nucleus.  For
each line with impact parameter $b$ the effective density profile
is given by
\begin{eqnarray}
L(b)&=&\sqrt{\pi} R\, e^{-\frac{b^2}{R^2}}\left[ 
1+c_0\ \left(\frac{1}{2}+\frac{b^2}{R^2}\right)\right. \nonumber \\
&+&\,c_1\left.
 \left(\frac{3}{4}+\frac{b^2}{R^2}+\left(\frac{b^2}{R^2}\right)^2
   \right)\right] \, , 
\end{eqnarray}
where $R$ is the radius of the nucleus and the density is 
parametrized as
\begin{equation}
\rho(r) =\rho(0)\, e^{-\frac{r^2}{R^2}}
 \left[1+c_0 \ \frac{r^2}{R^2}+c_1 \left(\frac{r^2}{R^2}\right)^2\right]
\end{equation}
given in the Landoldt-B\"ornstein Tables \cite{ref3}.

When the scattering is in both forward and backward directions the 
solution is known and given in eqs.~A25--A27 of ref.~\cite{ref2}.
They define the transition probabilities $f_{\pm}(\lambda)$ 
corresponding to the two directions. In terms of them we define
\begin{equation}
c = \frac{1}{3} -\frac{1}{2}\, \frac{f(5/6)}{f(1)} +
  \frac{1}{6}\, \frac{f(1/2)}{f(1)}
\end{equation}
and
\begin{equation}
d = \frac{1}{3}\, \left[1-\frac{f(1/2)}{f(1)}\right]
\end{equation}
and the matrices $M_{\pm}$ in analogy to eq.\ (6). Then
using differential cross sections we can obtain information on the 
corrections to the original angular distribution.  To be specific we
define differential cross sections
\begin{equation}
d\sigma(T^A, \hat{q})\quad{\rm and}\quad d\sigma(\hat{q})
\end{equation}
for the pions emerging from the nucleus and pions from the production
point, respectively.  The unit vector $\hat{q}$ denotes their angular direction.
Then
\begin{equation}
d\sigma(T\hat{q}) = M_+ d\sigma(\hat{q})+M_- dq(-\hat{q})
\end{equation}
since
\begin{equation}
M_+ +M_- = M\quad{\rm and}\quad f_+(\lambda)+f_-(\lambda)=f(\lambda)\, .
\end{equation} 
The above equations imply 
\begin{equation}
d\sigma(T,\hat{q})+d\sigma(T,-\hat{q})=
 M\left[d\sigma(\hat{q})+d\sigma(-\hat{q})\right]\, .
\end{equation}
A difference between forward and backward scattering in eq.~(8) comes
from the $h_{\pm}(W,\cos\theta))$ functions. In this article we
present results for the charge--exchange corrections integrated over
angles. It will be interesting to study in the future the modification
on the angular distribution caused by the nuclear rescattering \cite{ref4}.

This completes the formalism for calculating the transport matrix
in terms of the absorption and pion--nucleon cross sections;
all other parameters are defined by properties of the nuclei.

\section{Experimental Consequences and Comparisons}

There are two ways for using this model.  The first one uses
experimental data in order to determine the parameters.  The second
method is to use nuclear parameters from tables and calculate
the charge--exchange matrix.  Since the available data is very limited, one
uses the second method to compute the exchange matrix  and then compare
them with the few available experimental numbers.

\vspace*{-1.5cm}
\begin{figure}[htb]
\centering
\hspace*{-1.2cm}
\includegraphics[angle=0,width=8.5cm]{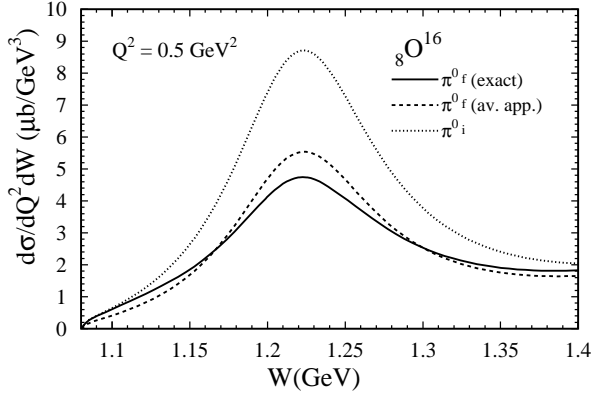}
\vspace*{-2.5cm}
\caption{\sf Electroproduction for $\pi^0$ on Oxygen target.}
\label{fig1}
\end{figure}

\begin{figure}[htb]
\vspace*{-2cm}
\hspace{-1.2cm}
\includegraphics[angle=0,width=8.5cm]{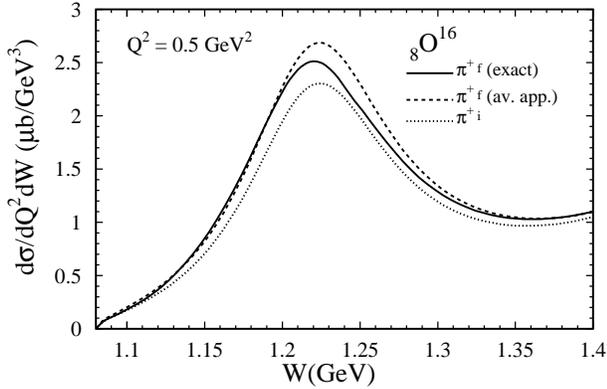}
\vspace*{-1.8cm}
\caption{\sf Electroproduction for $\pi^+$ on Oxygen target.}
\label{fig2}
\end{figure}

Before I present numerical values, I will state a general result
for medium heavy nuclei.  In a lepton nucleus interaction the 
pions which have the same charge as the exchanged current are reduced
substantially, by approximately 30-40\%, and for pions with different
charge there is a slight increase.  For instance, in electroproduction
and neutral current reactions the large reduction occurs for $\pi^0$'s,
while for charged--current neutrino--induced reactions the large 
correction occurs for $\pi^+$'s.  

Figures \ref{fig1} and \ref{fig2} show the electroproduction of 
pions on $_8{\rm O}^{16}$.  
The three curves were computed as follows: the dotted
line is the cross section without nuclear corrections (average over
protons and neutrons), the broken--line curve is the cross section 
within the averaging approximation \cite{ref2} 
and the solid curve is the cross section
including nuclear corrections without making any averaging, 
i.e., the exact transport problem.
We notice that in fig.\ \ref{fig1} 
the $\pi^0$ section is largely reduced (up to 40\%)
and in fig.\ \ref{fig2} the $\pi^+$ is increased by ~5\%.

\begin{figure}[htb]
\vspace*{-1cm}
\hspace*{-1.2cm}
\includegraphics[angle=0,width=8.5cm]{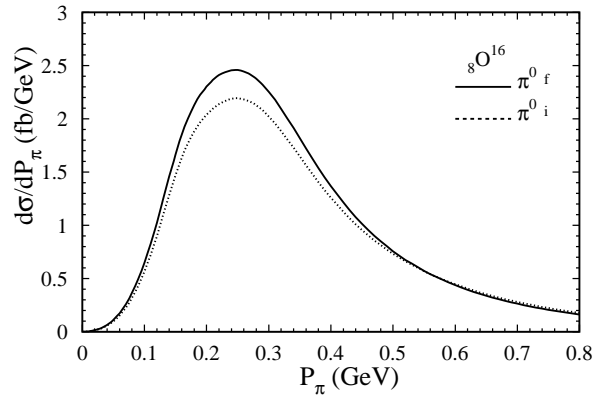}
\vspace*{-1.8cm}
\caption{\sf Charged--current neutrino production of $\pi^0$'s on Oxygen target. 
The solid line takes into account all nuclear corrections 
and dotted line represent without nuclear corrections.}
\label{fig3}
\end{figure}

\begin{figure}[htb]
\vspace*{-1cm}
\hspace{-1.cm}
\includegraphics[angle=0,width=8.5cm]{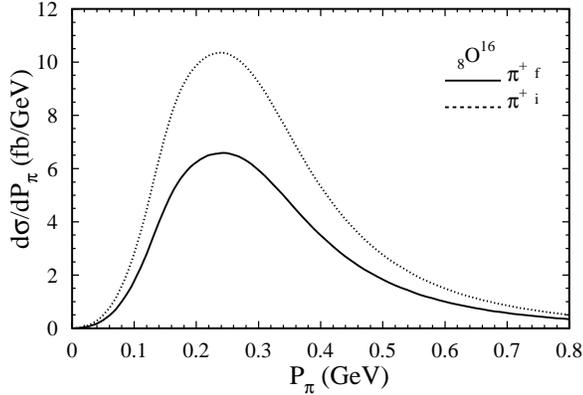}
\vspace*{-1.8cm}
\caption{\sf The same as in Fig. \ref{fig3} for $\pi^+$'s on Oxygen target.}
\label{fig4}
\end{figure}

Figures \ref{fig3} and \ref{fig4} 
show the same results for charged--current neutrino--production where
the $\pi^+$ is largely reduced and the $\pi^0$ yield slightly increased.
In Table \ref{tab2} 
we present the transport function $f(\lambda=1)$ for several
nuclei where the effect of the absorption cross section is now evident.\\

\renewcommand{\arraystretch}{1.2} 
\begin{center}
\begin{tabular}{|c|c|}
\hline
$_zT^A$ & $f(\lambda =1)$\\
\hline
$_6{\rm C}^{12}$ & 0.81\\
$_8{\rm O}^{16}$ & 0.80 \\
$_{18}{\rm A}^{40}$ & 0.65 \\
$_{26}{\rm Fe}^{56}$ & 0.63\\
\hline
\end{tabular}\\[2pt]
\end{center}
\begin{center}
Table 2: The transport function $f(\lambda=1)$.
\end{center}\label{tab2}

In addition, we computed the charge--exchange matrix for several values
of the absorption cross section and found that it affects $A$ by noticeable
amounts but the $3\times 3$ charge--exchange matrix very little (~3\%).
Values for the charge--exchange matrix were reported already and we give
values for several nuclei.\\

\noindent {\underline{Carbon}}
\begin{displaymath}
A=0.79\, , \quad M(_6{\rm C}^{12})=A
\left( \begin{array}{ccc} 0.83 & 0.14 & 0.04\\
 0.14 & 0.73 & 0.14 \\ 0.04 & 0.14 & 0.83
    \end{array}\right)
\end{displaymath}
\noindent {\underline{Oxygen}}
\begin{displaymath}
A=0.78\, , \quad M(_8{\rm O}^{16})=A
\left( \begin{array}{ccc} 0.78 & 0.16 & 0.06\\
  0.16 & 0.68 & 0.16\\ 0.06 & 0.16 & 0.78
   \end{array}\right)
\end{displaymath}
\noindent {\underline{Argon}}
\begin{displaymath}
A=0.66\, , \quad M(_{18}{\rm Ar}^{40}) = A
\left( 
\begin{array}{ccc} 
0.73 & 0.19 & 0.08\\
0.19 & 0.63 & 0.18\\ 
0.08 & 0.18 & 0.73
\end{array}\right)
\end{displaymath}
These matrices have been used for analyzing ratios of neutral to
charged currents \cite{ref5}.  We mention two examples in order
 to show the
changes that are introduced by rescatterings on nuclei.
\vspace{0.3cm}

\noindent {\bf\underline{Example 1}}  

This case considers the
ratio of neutral to charged--current reactions
\begin{equation}
R'(_zT^A) = \frac{\sigma(\nu_{\mu}+T\to \nu_{\mu}+\pi^0)}
   {2\sigma(\nu_{\mu}+T\to \mu^-+\pi^0)}
\end{equation}
which has been measured in two neutrino experiments 
\cite{ref6,ref7}.  The Columbia--BNL experiment had a complex
target consisting 75\% Al and 25\% C.  They reported the 
value \cite{ref6}
\begin{equation}
R'(_{13}{\rm Al}^{37}) = 0.17\pm 0.06\, .
\end{equation}
The Gargamelle experiment used a bubble chamber filled with 
Freon CF$_3$Br and reported \cite{ref7}
\begin{equation}
0.11 < R'({\rm Br}) < 0.22\, .
\end{equation}
The two experiments are consistent with each other.  The values
are, however, much smaller than the value calculated for the 
scattering on free protons and neutrons to be denoted by 
\begin{equation}
R(N_T) = 0.44\quad{\rm for}\quad\sin^2\theta_W=0.22\, .
\end{equation}
The agreement is restored \cite{ref8} when charge--exchange corrections 
are included
\begin{equation}
R_{\rm theory}(_{13}{\rm Al}^{37}) = 0.22\, .
\end{equation}
This example shows the important role that nuclear corrections
played in establishing the isospin content of the neutral current.
The same ratio could be, hopefully, measured in K2K, Superkamiokande
(SK) and Miniboone.  For the long--base--line experiment K2K  
there is an additional bonus:  the ratios should be larger in 
SK because as the $\nu_{\mu}$'s travel to Kamioka they oscillate
to $\nu_{\tau}$'s which do not contribute to the charged--current
cross section in the denominator.  The precise value of the 
ratio for the two detectors is calculable.
\vspace{0.3cm}

\noindent {\bf\underline{Example 2}}  

Another realistic case
occurred in the Gargamelle experiment where they measured the
production of pions in an enriched Freon target.  The average
energy of the neutrino beam was low, in the $\Delta$--resonance
region.  They observed the ratio \cite{ref7}:
\begin{equation}
\frac{\pi^+}{\pi^0} = 2.3\pm 0.9\, .
\end{equation}
The cross sections for $E_{\nu} = 2.0$ GeV in units of $10^{-38}{\rm cm}^2$
are
\begin{eqnarray}
\sigma(\nu p\to \mu^-p\pi^+) & = & 0.70\pm 0.10,\\
\sigma(\nu n\to \mu^-p\pi^0) & = & 0.20\pm 0.05,\\
\sigma(\nu n\to \mu^-n\pi^+) & = & 0.20 \pm 0.07\, .
\end{eqnarray}
It follows from these numbers that the ratio, averaged over
protons and neutrons, is
\begin{equation}
\left( \frac{\pi^+}{\pi^0}\right)_{{\rm free}\,\frac{(p+n)}{2}}
    =4.50\, ,
\end{equation}
which is far away from the value reported by Gargamelle. After
we include nuclear corrections for Bromine, the ratio becomes
\begin{equation}
\left( \frac{\pi^+}{\pi^0}\right)_{\rm nuclear\,\, corrected}
   = 2.64
\end{equation}
in good agreement with the experimental number.\\

Further information on the pion absorption cross section
could be extracted \cite{ref9}
from an Argonne experiment on Deuterium  \cite{ref10}
and data from the BEBC experiment \cite{ref11} on a Neon target.
Deuterium was considered to be an isoscalar
target of free protons and neutrons, and nuclear corrections were
introduced for Neon.  Unfortunately, the two experiments 
were at different energies.
However, it was possible to re-weight the data to the {\em same}
atmospheric neutrino energy spectrum \cite{ref12}
allowing for a direct comparison.
The parameters of the ANP model were found to be in good
qualitative agreement with the data \cite{ref9}.

The comparisons show that the qualitative properties of the pion
transport model are confirmed by experiments.  The remaining
question is to test the degree of its accuracy.  This can and
should be done in the running and upcoming experiments because
the model provides a means for studying the mixing and other
properties of neutrinos to a higher degree of accuracy.
Precise determinations of nuclear corrections are also important
for the discovery of CP--violation in the leptonic sector.\\

\noindent{\bf Acknowlegdement}

We wish to thank Dr.\ M.\ Sakuda for several useful discussions 
and the organizers for a productive and pleasant meeting at Gran
Sasso.

\end{document}